# Domain-adaptive Graph Attention-supervised Network for Cross-network Edge Classification

Xiao Shen, Mengqiu Shao, Shirui Pan, *Senior Member, IEEE*,
Laurence T. Yang, *Fellow, IEEE*, Xi Zhou

*Abstract*—Graph neural networks (GNNs) have shown great ability in modeling graphs, however, their performance would significantly degrade when there are noisy edges connecting nodes from different classes. To alleviate negative effect of noisy edges on neighborhood aggregation, some recent GNNs propose to predict the label agreement between node pairs within a single network. However, predicting the label agreement of edges across different networks has not been investigated yet. Our work makes the pioneering attempt to study a novel problem of cross-network homophilous and heterophilous edge classification (CNHHEC), and proposes a novel domain-adaptive graph attention-supervised network (DGASN) to effectively tackle the CNHHEC problem. Firstly, DGASN adopts multi-head GAT as the GNN encoder, which jointly trains node embeddings and edge embeddings via the node classification and edge classification losses. As a result, label-discriminative embeddings can be obtained to distinguish homophilous edges from heterophilous edges. In addition, DGASN applies direct supervision on graph attention learning based on the observed edge labels from the source network, thus lowering the negative effects of heterophilous edges while enlarging the positive effects of homophilous edges during neighborhood aggregation. To facilitate knowledge transfer across networks, DGASN employs adversarial domain adaptation to mitigate domain divergence. Extensive experiments on real-world benchmark datasets demonstrate that the proposed DGASN achieves the state-of-the-art performance in CNHHEC.

## I. INTRODUCTION

Graph Neural Networks (GNNs) [1] have made remarkable achievements on graph representation learning in various domains, such as social networks [2], protein-protein interaction networks [3], scene graphs [4], financial system [5], recommendation system [6], and healthcare [7]. The success of GNNs relies on a recursive neighborhood aggregation scheme, where the embedding of each node is updated by aggregating the embeddings of its neighbors. The neighborhood aggregation is based on the homophily assumption of networks [8] that suggests connected nodes generally have the same labels or similar features. With perfect homophily, aggregating the information from similar neighbors is indeed helpful for learning informative embeddings for various downstream tasks.

However, not all the information aggregated from the neighborhood is beneficial [9-12], since the real-world graphs usually contain structure noises, i.e., the noisy edges connecting nodes of different labels or features. Aggregating information through such noisy inter-class edges makes the information of different classes get mixed and consequently causes the over-smoothing issue [13], i.e., the embeddings of nodes from different classes become indistinguishable. Then, the performance of GNNs on downstream tasks can significantly degrade. To alleviate the negative effect of noisy edges, some recent GNNs [9-12, 14, 15] propose to predict the label agreement between node pairs, and then utilize the predicted output to filter out or down-weight the noisy inter-class edges during neighborhood aggregation. However, all these work [9-12, 14, 15] are conducted within a single network. To the best of our knowledge, predicting noisy edges

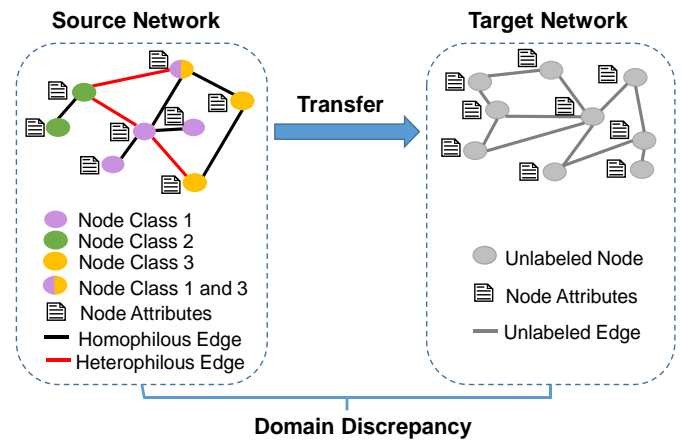

Fig. 1. An illustration of the CNHHEC problem. The source network has observed node and edge labels, while the target network is completely unlabeled. A node can be associated with multiple labels. The edges are labeled according to the label agreement between two nodes on each edge.

Manuscript submitted for review on December 19, 2022; revised 1 June 2023; accepted 24 August 2023. This work was supported in part by National Natural Science Foundation of China (No. 62102124, No. 62362020), Hainan Provincial Natural Science Foundation of China (No. 322RC570), the Specific Research Fund of The Innovation Platform for Academicians of Hainan Province (No. PT2300002941), and the Research Start-up Fund of Hainan University (No. KYQD(ZR)-22016). *(Corresponding author: Xiao Shen.)*

X. Shen and M. Shao are with the School of Computer Science and Technology, Hainan University, Haikou, China (e-mail: shenxiaocam@163.com; MengqiuShao@hainanu.edu.cn).

S. Pan is with the School of Information and Communication Technology, Griffith University, Gold Coast, Australia (e-mail: s.pan@griffith.edu.au).

L. T. Yang are with the School of Computer Science and Technology, Hainan University, Haikou, China, and the Department of Computer Science, St. Francis Xavier University, Canada (e-mail: ltyang@ieee.org).

X. Zhou is with the College of Tropical Crops, Hainan University, Haikou, China (e-mail: xzhou@hainanu.edu.cn).

across different networks has not been investigated yet.

To fill in this gap, in this work, we study a novel problem of cross-network homophilous and heterophilous edge classification (CNHHEC). Fig. 1 illustrates the CNHHEC problem, where each edge is labeled as either homophilous or heterophilous, according to the label agreement between two nodes on the edge. Specifically, a homophilous edge indicates that the two connected nodes share at least one common class-label. On the contrary, a heterophilous edge reflects that the two connected nodes have totally different class-labels. In CNHHEC, we have a fully labeled source network and a completely unlabeled target network, where the inherent domain discrepancy exists between the two networks. The goal is to accurately classify edges in the target network into either homophilous or heterophilous, by transferring the knowledge from the source network. For example, for cross-domain social recommendation, given a mature online social network (OSN) where many users have annotated tags indicating their interests, and a newly formed OSN where users are without labels, one can conduct CNHHEC to transfer the knowledge from the mature source network to predict homophilous and heterophilous edges in the new target network. Then, based on the predicted edge labels, one can recommend new connections (i.e. homophilous edges) between users who share similar interests, and also filter out noisy existing connections (i.e., heterophilous edges) between users with opposite hobbies.

In recent few years, cross-network node classification (CNNC) has gained increasing attention, which aims to transfer the knowledge from a fully labeled source network to accurately classify nodes in an unlabeled target network. The existing CNNC literatures [16-22] adopt a typical paradigm to integrate GNNs with domain adaptation to yield domain-adaptive GNNs. However, such domain-adaptive GNNs only focus on learning network-invariant node embeddings to address the downstream node classification task. While there is no existing domain-adaptive GNNs aiming to tackle the CNHHEC problem.

To fill in this gap, this work aims to propose the first domain-adaptive GNN to address the CNHHEC problem. In essence, we need to solve three challenges as follows: 1) Most existing GNNs focus on learning node embeddings, while how to learn informative edge embeddings to discriminate heterophilous edges from homophilous edges is not clear. 2) Due to message-passing among neighborhood, GNNs are vulnerable to noisy edges that connect nodes from different classes. How to alleviate negative effect of noisy heterophilous edges on neighborhood aggregation is a key issue we must address in the CNHHEC problem. 3) The source and target networks inherently have domain discrepancy, which hinders the GNN trained on the source to be directly applied to the target. How to reduce domain discrepancy across networks to learn network-invariant edge embeddings is an important issue should be tackled in CNHHEC.

To address the aforementioned challenges, we propose a novel domain-adaptive graph attention-supervised network (DGASN). Firstly, a multi-head graph attention network (GAT) [23] is adopted as the GNN encoder to learn node embeddings. Then, we construct edge embeddings based on the embeddings of two nodes on each edge. A key design of DGASN is to jointly train node embeddings and edge embeddings via both node classification and edge classification losses. As a result, not only label-discriminative node embeddings can be learned to separate nodes from different classes, but also label-discriminative edge embeddings can be learned to distinguish homophilous edges from heterophilous edges. Secondly, real-world graphs are usually noisy with connections between unrelated nodes, directly utilizing the adjacency matrix with fixed edge weights for neighborhood aggregation would inevitably introduce noises to node embeddings. Thus, instead of utilizing fixed edge weights, the proposed DGASN adopts GAT to automatically learn adaptive edge weights to capture various degree of importance of each neighbor, which can alleviate the negative effect of noisy heterophilous edges to some extent. However, in original GAT [23], the supervision on graph attention weights is limited and indirect [24, 25]. To improve the expressive power of graph attention weights, DGASN proposes to apply direct supervision on the attention weights learned by each GAT layer. Specifically, given the observed edge labels in the source network, a supervised attention loss is devised to guide the attention weights to be smaller for heterophilous edges while be larger for homophilous edges. As a result, the noisy heterophilous edges with smaller attention weights would have less effect during neighborhood aggregation. Thirdly, to learn network-invariant edge embeddings, DGASN adopts the representative adversarial domain adaptation method [26] to make the GNN encoder compete against the domain discriminator in an adversarial training manner. The proposed DGASN is trained in an end-to-end fashion to yield both label-discriminative and network-invariant edge embeddings to effectively address CNHHEC.

The contributions of this work are summarized as follows:

1) We are the first to formulate and study a novel problem of cross-network homophilous and heterophilous edge classification.

2) To effectively tackle CNHHEC, we propose a novel framework named DGASN, which jointly trains node embeddings and edge embeddings to distinguish heterophilous edges from homophilous edges, applies direct supervision on graph attention learning to lower negative effect of heterophilous edges, and learns network-invariant edge embeddings via adversarial domain adaptation.

3) Extensive experiments and ablation studies on benchmark datasets demonstrate the effectiveness of the proposed DGASN on the challenging CNHHEC problem.

## II. RELATED WORK

### A. Graph Neural Networks

The success of GNNs relies on recursive neighborhood aggregation, which iteratively updates the embedding of each central node by aggregating the embeddings among its neighborhood. Graph convolution network (GCN) [2] is the most representative GNN, which generalizes convolution operation in computer vision to graph data. Inspired by GCN, many GNNs [3, 27, 28] have been proposed. Instead of treating all neighbors equally, the attention-based GNNs propose to learn adaptive attention weights to capture various degree of importance of each neighbor to represent the center node. GAT [23] is the most representative attention-based GNN, which adopts a masked self-attention mechanism to learn adaptive edge weights in neighborhood aggregation. Inspired by GAT, various attention-based GNNs [29-31] have been proposed. However, these attention-based GNNs lack direct supervision on the learning of attention weights, which easily causes over-fitting and over-smoothing [25], especially when graphs are noisy. To remedy this limitation, SuperGAT [25] employs a self-supervised task of link prediction to directly supervise attention weights. Inspired by SuperGAT, the proposed DGASN also applies direct supervision on graph attention learning so as to improve the expressive power of attention weights. However, our work is inherently different from SuperGAT in that instead of supervising the attention weights to distinguish connected nodes (i.e. edges) from disconnected nodes (i.e. non-edges) within a single network, our DGASN supervises the attention weights to distinguish homophilous edges from heterophilous edges across different networks with distribution shifts.

Real-world graphs usually contain structure noises, and aggregating neighborhood information through such noisy inter-class edges would be harmful to the downstream tasks [9-12, 14, 15]. To address this, some GNNs propose to predict the label agreement between node pairs. LAGCN [11] employs a label-aware edge classifier to filter out existing neighbors with different labels and add new edges between disconnected but the same labeled nodes. GAM [10] proposes an auxiliary model to predict the probability of two nodes sharing the same label, and used the predicted output to regularize the node classification model. EGAI [9] proposes to remove inter-class edges so as to achieve a high-quality neighborhood aggregation. NRGNN [14] adopts a GNN-based edge predictor to link unlabeled nodes with similar labeled nodes. RS-GNN [12] employs a link predictor to down-weight noisy edges and densify graph by adding new edges that connect nodes with high similarity. All these work focus on predicting the label agreement between node pairs within a single network. In contrast, our work aims to predict the label agreement of edges across different networks.

### B. Cross-network Node Classification

Motivated by the significant achievements of knowledge transfer [32-34] in Computer Vision (CV) and Natural Language Processing (NLP), recently, a line of work proposes to transfer the knowledge across graph-structured data. Cross-network node classification (CNNC) [16-22] aims to transfer the node classification knowledge from a labeled source network to classify unlabeled nodes in a target network. CDNE [16] is a pioneering CNNC algorithm, which employs two stacked auto-encoders (SAEs) to reconstruct the topological proximity matrix of the source and target networks respectively. Then, it minimizes MMD and class-conditional MMD [35] to learn network-invariant node representations. AdaGCN [17] leverages GCN [2] to learn node representations and employs the Wasserstein distance guided adversarial domain adaptation [36] to mitigate domain discrepancy. ACDNE [19] employs dual feature extractors with different learnable parameters to separately learn node representations from neighbor representations so as to jointly capture homophily and heterophily between nodes. UDAGCN [20] employs dual GCN to jointly capture local and global consistency for neighborhood aggregation. Similar to UDAGCN, ASN [18] also employs a dual GCN for graph representation learning, and further adopts Domain Separation Networks (DSN) [37] to separate domain-private and domain-shared representations.

The most representative adversarial domain adaptation method, which inserts a gradient reversal layer (GRL) [26] between domain discriminator and generator, has been adopted in [18-20] to guide network-invariant node representations.

Our work is related to the recent CNNC algorithms [16-20], since we also integrate GNN with domain adaptation to yield domain-adaptive GNN. However, our work is inherently different from the CNNC literatures since we study a new problem of CNHHEC.

### C. Cross-network Edge Classification

The cross-network edge classification problem has been studied in some early literatures. Tang *et al.* [38, 39] proposed a TranFG model to classify social relationships in a target network by borrowing the knowledge from a source network. Qi *et al.* [40] proposed a cross-network link prediction model to predict unseen links in the target network by transferring the link information from the source network. Shen *et al.* [41, 42] proposed a cross-network learning model to predict inactive edges for influence maximization in the target network by leveraging the knowledge learned from the source network.

Our work is inherently different from the early cross-network edge classification literatures [38-42] in three aspects. Firstly, the early literatures adopted the feature engineering approach to manually define explicit edge features based on social theories [38, 39] or topological features [40-42]. Instead, our DGASN employs the attention-supervised GNN to automatically learn latent edge embeddings in an end-to-end manner. Secondly, previous cross-network edge classification algorithms adopted the early domain adaptation methods, such as re-sampling [40] or self-training [41, 42]. While our DGASN adopts more powerful adversarial domain adaptation method. Thirdly, the definition of edge labels is different between our work and previous literatures. Specifically, [38, 39] define edge labels as the type of social relationships, [40] defines edge labels as links or non-links, [41, 42] defines edge labels as active or inactive for influence maximization. In contrast, we define edge labels as homophilous or heterophilous, according to the label agreement between two connected nodes.

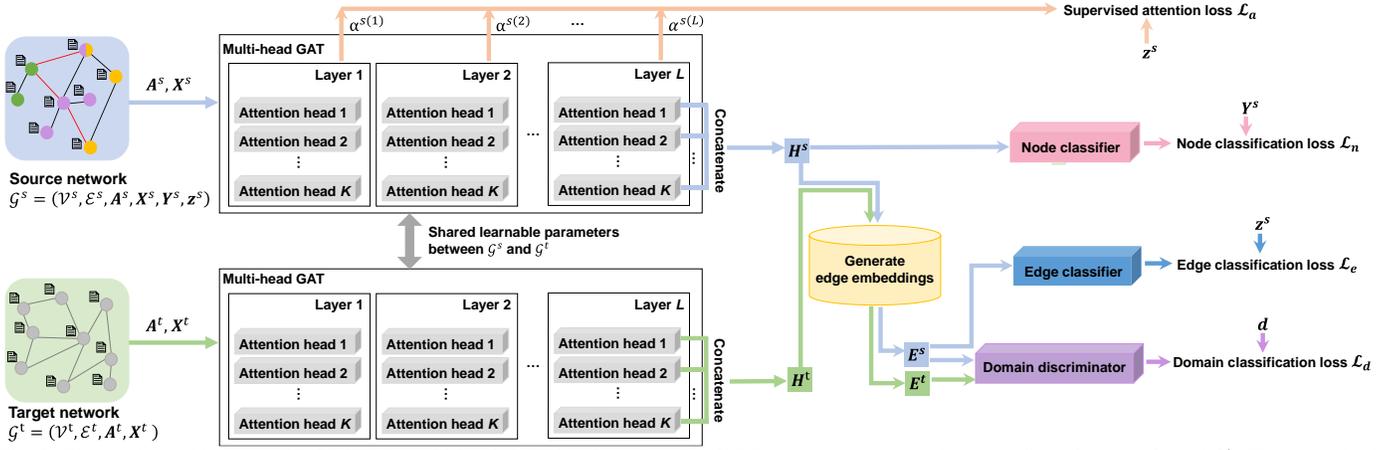

Fig. 2. The model architecture of DGASN. A multi-head GAT is adopted as the GNN encoder to learn node embeddings for both $\mathcal{G}^s$ and $\mathcal{G}^t$. The supervised attention loss is applied on the source attention weights at each layer of GAT, supervised by the observed edge labels in $\mathcal{G}^s$. The node classification loss and edge classification loss are guided by the observed node and edge labels in $\mathcal{G}^s$. The domain classification loss is guided by the domain labels of edges.

## III. PROPOSED MODEL

In this section, we firstly formulate the CNHHEC problem, and then elaborate on the proposed DGASN. Fig. 2 shows the model architecture of DGASN, which contains a multi-head GAT encoder, a node classifier, an edge classifier, and a domain discriminator.

### A. Problem Definition

Let $\mathcal{G} = (\mathcal{V}, \mathcal{E}, A, X, Y, z)$ denote an undirected network with a set of nodes $\mathcal{V}$, a set of undirected edges $\mathcal{E}$, an adjacency matrix $A \in \mathbb{R}^{|\mathcal{V}| \times |\mathcal{V}|}$, a node attribute matrix $X \in \mathbb{R}^{|\mathcal{V}| \times \mathcal{W}}$, a node label matrix $Y \in \mathbb{R}^{|\mathcal{V}| \times \mathcal{C}_\mathcal{V}}$, and an edge label vector $z \in \mathbb{R}^{|\mathcal{E}|}$, where $|\mathcal{V}|, |\mathcal{E}|, \mathcal{W}, \mathcal{C}_\mathcal{V}$ denote the number of nodes, edges, node attributes, and node label categories in $\mathcal{G}$ respectively. Specifically, if there is an edge connecting $v_i$ and $v_j$, i.e., $(v_i, v_j) \in \mathcal{E}$, then $A_{ij} = A_{ji} = 1$; otherwise, i.e., $(v_i, v_j) \notin \mathcal{E}$, $A_{ij} = A_{ji} = 0$. $X_{ik} = 1$ if node $v_i$ is associated with the $k$-th node attribute, otherwise $X_{ik} = 0$. $Y_{ic} = 1$ if node $v_i$ is associated with node label $c$, otherwise $Y_{ic} = 0$. Note that for multi-label node classification, a node can be associated with multiple labels. $z_{(v_i, v_j)}$ is the edge label of $(v_i, v_j)$, where $z_{(v_i, v_j)} = 1$ if $(v_i, v_j)$ is a homophilous edge and $z_{(v_i, v_j)} = 0$ if $(v_i, v_j)$ is a heterophilous edge. Specifically, $z_{(v_i, v_j)} = 1$ if $v_i$ and $v_j$ share at least one common node label, i.e., $\exists c \in \{1, 2, \cdots, \mathcal{C}_\mathcal{V}\}, Y_{ic} = Y_{jc} = 1$. On the contrary, if $v_i$ and $v_j$ have totally different node labels, then $z_{(v_i, v_j)} = 0$.

In the CNHHEC problem, we have a fully labeled source network $\mathcal{G}^s = (\mathcal{V}^s, \mathcal{E}^s, A^s, X^s, Y^s, z^s)$ where all nodes and edges have observed labels, and a completely unlabeled target network $\mathcal{G}^t = (\mathcal{V}^t, \mathcal{E}^t, A^t, X^t)$ where all nodes and edges do not have labels. Note that $\mathcal{G}^s$ and $\mathcal{G}^t$ have different distributions of network topology and node attributes, while $\mathcal{G}^s$ and $\mathcal{G}^t$ share the common label space. The goal of CNHHEC is to take advantage of the fully labeled source data and unlabeled target data to accurately predict edge labels of $\mathcal{G}^t$ in an end-to-end manner. For clarity, the frequently used notations are summarized in Table I.

### B. Joint Training of Node and Edge Embeddings

The GCN-like models [2, 3, 27, 28] are vulnerable to noisy inter-class edges, since all neighbors are treated equally with the fixed edge weights during neighborhood aggregation. While GAT [23] can alleviate negative effects of noisy edges, by automatically assigning adaptive attention edge weights to different neighbors to reflect their important degree to the target node. Thus, in the proposed DGASN, we opt for GAT as the GNN encoder for node embedding learning.

#### 1) Node Embeddings and Node Classification

Each $l$-th GAT layer learns the node embedding of $v_i$ by adaptively aggregating the embeddings of itself and its neighbors at previous layer and then applies an ELU nonlinear activation function, which is expressed as:

$$h_i^{(l)} = \text{ELU}\left(\sum_{j \in \mathcal{N}_i \cup \{i\}} \hat{\alpha}_{(v_i, v_j)}^{(l)} W^{(l)} h_j^{(l-1)}\right), \forall 1 \leq l \leq L \quad (1)$$

where $h_i^{(l)} \in \mathbb{R}^{\mathbb{d}^{(l)}}$ is the node embedding of $v_i$ at the $l$-th GAT layer, $\mathbb{d}^{(l)}$ is the number of embedding dimensions at the $l$-th

TABLE I
FREQUENTLY USED NOTATIONS.

| Notations | Descriptions |
|---|---|
| $\mathcal{G}$ | A network |
| $\mathcal{G}^s, \mathcal{G}^t$ | Source network and target network |
| $|\mathcal{V}|, |\mathcal{E}|$ | Number of nodes and edges in $\mathcal{G}$ |
| $\mathcal{W}$ | Number of node attributes in $\mathcal{G}$ |
| $v_i$ | $i$-th node in $\mathcal{G}$ |
| $x_i$ | Node attribute vector of $v_i$ |
| $y_i$ | Node label vector of $v_i$ |
| $h_i$ | Node embedding vector of $v_i$ |
| $\mathbb{d}^{(l)}$ | Node embedding dimensions at $l$-th GAT layer |
| $(v_i, v_j)$ | An edge connecting $v_i$ and $v_j$ in $\mathcal{G}$ |
| $z_{(v_i, v_j)}$ | Edge label of $(v_i, v_j)$ |
| $e_{(v_i, v_j)}$ | Edge embedding vector of $(v_i, v_j)$ |
| $\alpha_{(v_i, v_j)}^{(l)}$ | Attention edge weight of $(v_i, v_j)$ at $l$-th GAT layer |
| $K$ | Number of attention heads |
| $L$ | Number of GAT layers |
| $f_h, f_y, f_z, f_d$ | GNN encoder, node classifier, edge classifier and domain discriminator |
| $\theta_h, \theta_y, \theta_z, \theta_d$ | Learnable parameters of $f_h, f_y, f_z, f_d$ |

GAT layer, $h_i^{(0)} = x_i \in \mathbb{R}^\mathcal{W}$ is the input node attribute vector of $v_i$, $W^{(l)} \in \mathbb{R}^{\mathbb{d}^{(l)} \times \mathbb{d}^{(l-1)}}$ is the learnable weight matrix of the $l$-th layer, and $L$ is the number of GAT layers. $\mathcal{N}_i = \{j | A_{ij} = 1\}$ denotes the first-order neighbors of $v_i$.

The key idea of GAT is to automatically learn adaptive edge weight $\hat{\alpha}_{(v_i,v_j)}^{(l)}$ by a self-attention mechanism. Specifically, by taking the embeddings of $v_i$ and $v_j$ at $(l-1)$-th layer as the inputs, the $l$-th GAT layer learns adaptive edge weight of $(v_i, v_j)$, as:

$$\alpha_{(v_i,v_j)}^{(l)} = \text{LeakyReLU}\left(a^{(l)^T}[W^{(l)} h_i^{(l-1)} \| W^{(l)} h_j^{(l-1)}]\right) \quad (2)$$

where $a^{(l)} \in \mathbb{R}^{2\mathbb{d}^{(l)}}$ is a learnable vector of the $l$-th layer, $\cdot^T$ is the transposition operation, and $[\cdot \| \cdot]$ is the concatenation operation. To take network topology into account in graph attention learning, a masked attention mechanism is adopted to only compute $\alpha_{(v_i,v_j)}^{(l)}$ for $j \in \mathcal{N}_i \cup \{i\}$. In addition, to make the attention edge weights in Eq. (2) easily comparable across different nodes, $\alpha_{(v_i,v_j)}^{(l)}$ is normalized across all choices of $j$ via the Softmax function, as:

$$\hat{\alpha}_{(v_i,v_j)}^{(l)} = \frac{\exp\left(\alpha_{(v_i,v_j)}^{(l)}\right)}{\sum_{k \in \mathcal{N}_i \cup \{i\}} \exp\left(\alpha_{(v_i,v_k)}^{(l)}\right)} \quad (3)$$

where $\hat{\alpha}_{(v_i,v_j)}^{(l)}$ is the relative importance degree of $v_j$ among the neighborhood of $v_i$ (including $v_i$ itself) at the $l$-th GAT layer.

To make the self-attention mechanism more stable, we employ the multi-head attention mechanism to adopt $K$ independent attention heads to learn node embedding of $v_i$ in Eq. (1), and then concatenate the output embeddings of all attention heads as the node embedding of $v_i$. Then, the number of node embedding dimensions at the $l$-th GAT layer would become $K\mathbb{d}^{(l)}$.

For simplicity of notation, we denote the aforementioned multi-layer and multi-head GAT encoder as $f_h(\cdot; \theta_h)$, where $\theta_h$ represents the trainable parameters of node embedding learning. In the context of cross-network classification, we employ the shared learnable parameters $\theta_h$ between $\mathcal{G}^s$ and $\mathcal{G}^t$ to generate the cross-network node embeddings, as:

$$H^s = \{h_i^s\}_{i=1}^{|\mathcal{V}^s|} = f_h(A^s, X^s; \theta_h)$$
$$H^t = \{h_j^t\}_{j=1}^{|\mathcal{V}^t|} = f_h(A^t, X^t; \theta_h) \quad (4)$$

where $h_i^s, h_j^t \in \mathbb{R}^{K\mathbb{d}^{(L)}}$ are the final node embedding vectors of $v_i^s$ and $v_j^t$ learned by the deepest layer of multi-head GAT.

Then, a node classifier $f_y(\cdot; \theta_y)$ parameterized by $\theta_y$ constructed by a multi-layer perceptron (MLP) is added on the final node embedding vector of each node $v_i \in \mathcal{V}^s \cup \mathcal{V}^t$:

$$\hat{y}_i = f_y(h_i; \theta_y) \quad (5)$$

where $\hat{y}_i \in \mathbb{R}^{\mathcal{C}_\mathcal{V}}$ is the predicted node label vector of $v_i$ over $\mathcal{C}_\mathcal{V}$ categories. Given the observed node labels in $\mathcal{G}^s$, the sigmoid cross-entropy loss is adopted to define the multi-label node classification loss, as:

$$\mathcal{L}_n = -\frac{1}{|\mathcal{V}^s|} \Sigma_{v_i \in \mathcal{V}^s} \Sigma_{c=1}^{\mathcal{C}_\mathcal{V}} \begin{pmatrix} Y_{ic}^s \log \hat{Y}_{ic}^s + \\ (1 - Y_{ic}^s) \log(1 - \hat{Y}_{ic}^s) \end{pmatrix} \quad (6)$$

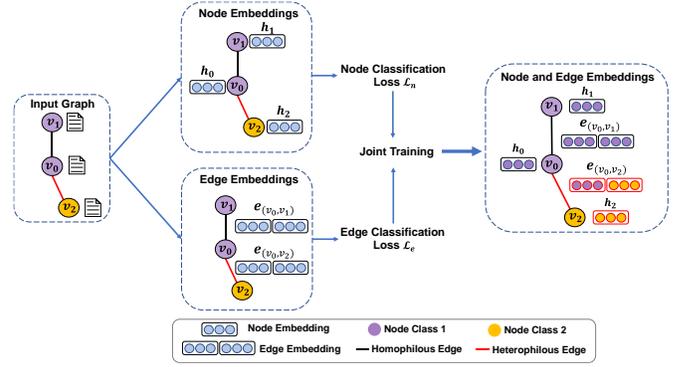

Fig. 3. An illustration of the joint training of node and edge embeddings via optimizing both node classification loss $\mathcal{L}_n$ and edge classification loss $\mathcal{L}_e$. Edge embedding is constructed based on the embeddings of two nodes on the edge. Label-discriminative edge embeddings can be implicitly learned by optimizing $\mathcal{L}_n$ and explicitly learned by optimizing $\mathcal{L}_e$.

where $Y_{ic}^s$ is the ground-truth node label of $v_i^s$, $Y_{ic}^s = 1$ if $v_i^s$ is associated with label $c$; otherwise, $Y_{ic}^s = 0$. $\hat{Y}_{ic}^s$ is the predicted probability of $v_i^s$ belonging to category $c$. Minimizing $\mathcal{L}_n$ guides label-discriminative node embeddings, which are conducive to distinguish homophilous edges (connecting nodes of the same label) from heterophilous edges (connecting nodes with different labels).

*2) Edge Embeddings and Edge Classification*

Next, we generate the edge embeddings based on the embeddings of two nodes on each edge. Five operators can be adopted to construct edge embeddings, following [43]:

$$\text{Concatenate: } e_{(v_i,v_j)} = [h_i \| h_j]$$
$$\text{Hadamard: } e_{(v_i,v_j)} = h_i \odot h_j$$
$$\text{Average: } e_{(v_i,v_j)} = (h_i + h_j)/2$$
$$\text{L1: } e_{(v_i,v_j)} = |h_i - h_j|$$
$$\text{L2: } e_{(v_i,v_j)} = |h_i - h_j|^2 \quad (7)$$

where $e_{(v_i,v_j)}$ denotes the edge embedding vector of $(v_i, v_j)$, and $\odot$ denotes the element-wise Hadamard product operator. In the proposed DGASN, we opt for the Concatenate operator to construct edge embeddings.

An edge classifier $f_z(\cdot; \theta_z)$ parameterized by $\theta_z$ is constructed by an MLP. Given the edge embedding $e_{(v_i,v_j)}$ as the input, the edge classifier outputs:

$$\hat{z}_{(v_i,v_j)} = f_z\left(e_{(v_i,v_j)}; \theta_z\right) \quad (8)$$

where $\hat{z}_{(v_i,v_j)}$ is the predicted probability of $(v_i, v_j)$ to be homophilous. Given observed edge labels of $\mathcal{G}^s$, the edge classification loss is defined as:

$$\mathcal{L}_e = -\frac{1}{|\mathcal{E}^s|} \Sigma_{(v_i,v_j) \in \mathcal{E}^s} \begin{pmatrix} z_{(v_i,v_j)}^s \log \hat{z}_{(v_i,v_j)}^s + \\ (1 - z_{(v_i,v_j)}^s) \log(1 - \hat{z}_{(v_i,v_j)}^s) \end{pmatrix} \quad (9)$$

where $z_{(v_i,v_j)}^s$ is the ground-truth edge label of $(v_i, v_j) \in \mathcal{E}^s$, $z_{(v_i,v_j)}^s = 1$ if $(v_i, v_j)$ is homophilous and $z_{(v_i,v_j)}^s = 0$ if $(v_i, v_j)$ is heterophilous.

Fig. 3 illustrates the joint training process of node and edge embeddings in the proposed DGASN, where the node and edge embeddings are learned end-to-end by optimizing both node classification loss $\mathcal{L}_n$ and edge classification loss $\mathcal{L}_e$ together.

It is worth noting that the edge embedding $e_{(v_i,v_j)}$ is constructed based on the embeddings of two nodes $h_i$ and $h_j$ connected by each edge, via various operators in Eq. (7). On one hand, by optimizing the node classification loss $\mathcal{L}_n$, label-discriminative node embeddings separating different node classes can be learned. Then, a homophilous edge (e.g. $e_{(v_0,v_1)}$ in Fig. 3) connecting two nodes from the same class would construct an edge embedding given two similar node embeddings as the inputs, whereas, a heterophilous edge (e.g. $e_{(v_0,v_2)}$ in Fig. 3) connecting two nodes from different classes would construct an edge embedding given two dissimilar node embeddings as the inputs. As a result, label-discriminative edge embeddings to distinguish homophilous edges from heterophilous edges can be implicitly learned upon such label-discriminate node embeddings. On the other hand, by directly optimizing the edge classification loss $\mathcal{L}_e$ supervised by the ground-truth edge labels, label-discriminative edge embeddings can be explicitly learned. Since the node and edge embeddings are jointly learned end-to-end, such explicitly label-discriminative edge embeddings can in turn yield more label-discriminative node embeddings. The ablation study in Section IV. C verifies that both node classification loss $\mathcal{L}_n$ and edge classification loss $\mathcal{L}_e$ are indispensable in the proposed DGASN to tackle the CNHHEC problem.

### C. Direct Supervision on Graph Attention Learning

GAT easily causes overfitting in graph attention learning, due to the limited and indirect supervision on learnable attention parameters [25]. In original GAT [23], the supervision on attention weights only comes from the node classification loss. In addition, it has been theoretically and empirically shown that if there are noisy edges connecting nodes from different classes, then by increasing the depth of GAT, the over-smoothing issue easily arises, where the embeddings of nodes from different classes become indistinguishable [24, 25]. The over-smoothing of node embeddings would severely impede the identification of heterophilous edges which connect nodes from different classes.

To go beyond the limits of original GAT, the proposed DGASN gives more supervision on the learnable parameters of graph attention, where the supervised signals come from not only the node classification loss, but also the edge classification loss, the supervised attention loss and the domain classification loss (will be introduced later). Actually, the ideal attention mechanism should assign larger weights to the intra-class neighbors linked by homophilous edges, while smaller weights to the inter-class neighbors linked by heterophilous edges. To this end, the proposed DGASN applies direct supervision on graph attention learning, according to the observed edge labels in $\mathcal{G}^s$.

Fig. 4 illustrates the idea of direct attention supervision on $\mathcal{G}^s$ at each $l$-th GAT layer. Firstly, for each source edge $(v_i, v_j) \in \mathcal{E}^s$, we compute the attention weight by averaging over $K$ attention heads, i.e., $\alpha^{s(l)}_{(v_i,v_j)} = \frac{1}{K}\sum_{k=1}^{K} \alpha^{s(l)(k)}_{(v_i,v_j)}$, where $\alpha^{s(l)(k)}_{(v_i,v_j)}$ is the unnormalized attention edge weight of $(v_i,v_j)$ in

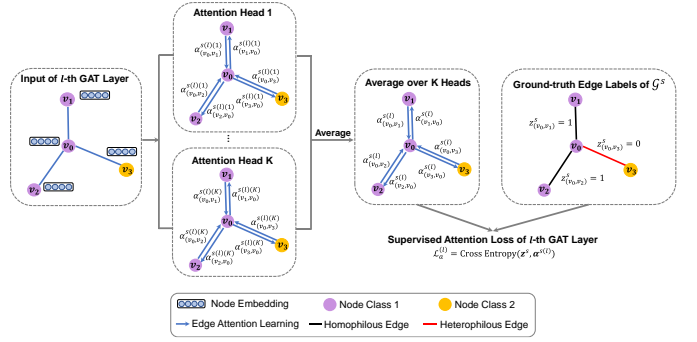

Fig. 4. An illustration of direct supervision on graph attention learning on $\mathcal{G}^s$ at the $l$-th GAT layer. Firstly, asymmetric attention weights are learned for each undirected edge by averaging over multiple attention heads. Then, given the ground-truth edge labels of $\mathcal{G}^s$, the supervised attention loss is devised to guide larger (lower) asymmetric attention weights for each homophilous (heterophilous) edge.

Eq. (2) learned by the $k$-th head at the $l$-th GAT layer. In addition, note that for an undirected edge $(v_i, v_j)$, two asymmetric attention weights $\alpha^{s(l)}_{(v_i,v_j)}$ and $\alpha^{s(l)}_{(v_j,v_i)}$ are learned by GAT, i.e., $\alpha^{s(l)}_{(v_i,v_j)} \neq \alpha^{s(l)}_{(v_j,v_i)}$, where $\alpha^{s(l)}_{(v_i,v_j)}$ reflects the importance of $v_j$ to $v_i$ during the node embedding learning of $v_i$, while $\alpha^{s(l)}_{(v_j,v_i)}$ indicates the importance of $v_i$ to $v_j$ when learning the node embedding of $v_j$. Given the ground-truth edge labels in $\mathcal{G}^s$, we devise the supervised attention loss to guide larger (lower) asymmetric attention weights on each homophilous (heterophilous) edge. Accordingly, the supervised attention loss at each $l$-th GAT layer is defined as:

$$\mathcal{L}_a^{(l)} = -\frac{1}{2|\mathcal{E}^s|}\sum_{(v_i,v_j)\in\mathcal{E}^s}\begin{pmatrix} z^s_{(v_i,v_j)}\left(\log\sigma\left(\alpha^{s(l)}_{(v_i,v_j)}\right) + \log\sigma\left(\alpha^{s(l)}_{(v_j,v_i)}\right)\right) \\ +\left(1-z^s_{(v_i,v_j)}\right)\begin{pmatrix}\log\left(1-\sigma\left(\alpha^{s(l)}_{(v_i,v_j)}\right)\right) \\ +\log\left(1-\sigma\left(\alpha^{s(l)}_{(v_j,v_i)}\right)\right)\end{pmatrix} \end{pmatrix} \quad (10)$$

where $\sigma$ is a sigmoid activation function. Minimizing Eq. (10) guides both $\sigma\left(\alpha^{s(l)}_{(v_i,v_j)}\right)$ and $\sigma\left(\alpha^{s(l)}_{(v_j,v_i)}\right)$ to be 1, if $(v_i,v_j)$ is a homophilous edge, i.e., $z^s_{(v_i,v_j)}=1$; in contrast, both $\sigma\left(\alpha^{s(l)}_{(v_i,v_j)}\right)$ and $\sigma\left(\alpha^{s(l)}_{(v_j,v_i)}\right)$ would be optimized to 0 for each heterophilous edge, i.e., $z^s_{(v_i,v_j)}=0$.

In addition, note that the number of homophilous edges is much larger than that of heterophilous edges in networks with homophily. Thus, directly minimizing Eq. (10) makes the supervised attention learning bias towards the homophilous edges. To address this, we incorporate a cost-sensitive parameter $\gamma > 1$ to modify Eq. (10) to impose the supervised attention learning focus more on the scarce heterophilous edges, which is expressed as:

$$\mathcal{L}_a^{(l)} = -\frac{1}{2|\mathcal{E}^s|}\sum_{(v_i,v_j)\in\mathcal{E}^s}\begin{pmatrix} z^s_{(v_i,v_j)}\left(\log\sigma\left(\alpha^{s(l)}_{(v_i,v_j)}\right) + \log\sigma\left(\alpha^{s(l)}_{(v_j,v_i)}\right)\right) \\ +\gamma\left(1-z^s_{(v_i,v_j)}\right)\begin{pmatrix}\log\left(1-\sigma\left(\alpha^{s(l)}_{(v_i,v_j)}\right)\right) \\ +\log\left(1-\sigma\left(\alpha^{s(l)}_{(v_j,v_i)}\right)\right)\end{pmatrix} \end{pmatrix} \quad (11)$$

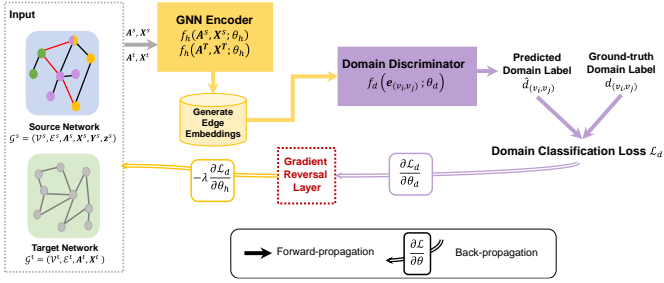

Fig. 5. An illustration of the adversarial domain adaptation. The domain discriminator $f_d$ and the GNN encoder $f_h$ are trained in an adversarial manner, by minimizing and maximizing the domain classification loss $\mathcal{L}_d$ respectively. A gradient reversal layer (GRL) is inserted between $f_d$ and $f_h$ during back-propagation to reverse the partial derivative of the domain classification loss w.r.t. the GNN encoder (i.e. $\frac{\partial \mathcal{L}_d}{\partial \theta_h}$) by multiplying it by $-\lambda$. Then, $\max_{\theta_h}\{\mathcal{L}_d\}$ and $\min_{\theta_d}\{\mathcal{L}_d\}$ can be simultaneously updated during each back-propagation.

By applying Eq. (11) on all the $L$ GAT layers, the total supervised attention loss is defined as:

$$\mathcal{L}_a = \sum_{l=1}^{L} \mathcal{L}_a^{(l)} \quad (12)$$

Minimizing $\mathcal{L}_a$ directly supervises GAT to learn more expressive attention weights, i.e., larger weights for homophilous edges while smaller weights for heterophilous edges.

### D. Adversarial Domain Adaptation

The source and the target networks inherently have different data distributions, which hinders the edge classifier $f_z$ trained on the source to be directly applied to the target. To mitigate the domain discrepancy between $\mathcal{G}^s$ and $\mathcal{G}^t$, the proposed DGASN employs the adversarial domain adaptation method [26] by training the GNN encoder $f_h$ and the domain discriminator $f_d$ in an adversarial manner. The domain discriminator $f_d(\cdot;\theta_d)$ parameterized by $\theta_d$ is constructed by an MLP, taking the edge embedding $\boldsymbol{e}_{(v_i,v_j)}$ as the input, it outputs:

$$\hat{d}_{(v_i,v_j)} = f_d\left(\boldsymbol{e}_{(v_i,v_j)};\theta_d\right) \quad (13)$$

where $\hat{d}_{(v_i,v_j)}$ is the predicted probability of $(v_i,v_j)$ coming from $\mathcal{G}^t$. Then, given the ground-truth domain labels of all edges in $\mathcal{G}^s$ and $\mathcal{G}^t$, the domain classification loss is defined as:

$$\mathcal{L}_d = -\frac{1}{|\mathcal{E}^s|+|\mathcal{E}^t|}\sum_{(v_i,v_j)\in\mathcal{E}^s\cup\mathcal{E}^t}\begin{pmatrix}d_{(v_i,v_j)}\log\hat{d}_{(v_i,v_j)} + \\ \left(1-d_{(v_i,v_j)}\right)\log\left(1-\hat{d}_{(v_i,v_j)}\right)\end{pmatrix} \quad (14)$$

where $d_{(v_i,v_j)}$ is the ground-truth domain label of $(v_i,v_j)$, $d_{(v_i,v_j)} = 1$ if $(v_i,v_j) \in \mathcal{E}^t$ and $d_{(v_i,v_j)} = 0$ if $(v_i,v_j) \in \mathcal{E}^s$.

In adversarial domain adaptation, the domain discriminator $f_d$ and the GNN encoder $f_h$ act as two players in a min-max game. On one hand, $\min_{\theta_d}\{\mathcal{L}_d\}$ guides $f_d$ to accurately predict which network an edge comes from. On the other hand, $\max_{\theta_h}\{\mathcal{L}_d\}$ guides $f_h$ to learn network-invariant edge embeddings to deceive $f_d$. That is, $f_d$ and $f_h$ are competing against each other by minimizing and maximizing the domain classification loss $\mathcal{L}_d$ in an adversarial way. In order to simultaneously update the learnable parameters of $f_d$ and $f_h$, we follow [26] to insert a GRL between $f_d$ and $f_h$ during

**Algorithm 1: DGASN**

**Input**: Fully labeled Source network $\mathcal{G}^s = (\mathcal{V}^s, \mathcal{E}^s, \boldsymbol{A}^s, \boldsymbol{X}^s, \boldsymbol{Y}^s, \boldsymbol{z}^s)$, completely unlabeled target network $\mathcal{G}^t = (\mathcal{V}^t, \mathcal{E}^t, \boldsymbol{A}^t, \boldsymbol{X}^t)$.

1. Initialize learnable parameters $\theta_h, \theta_y, \theta_z, \theta_d$;
2. **while** not max epoch **do**:
3.     Learn cross-network node embeddings $\{\boldsymbol{h}_i^s\}_{i=1}^{|\mathcal{V}^s|}$ and $\{\boldsymbol{h}_j^t\}_{j=1}^{|\mathcal{V}^t|}$ by GNN encoder $f_h$ in Eq. (4);
4.     Compute node classification loss $\mathcal{L}_n$ based on $\{(\boldsymbol{h}_i^s, \boldsymbol{y}_i^s)\}_{i=1}^{|\mathcal{V}^s|}$ in Eq. (6);
5.     Generate cross-network edge embeddings $\{\boldsymbol{e}_{(v_i,v_j)}^s\}_{(v_i,v_j)\in\mathcal{E}^s}$ and $\{\boldsymbol{e}_{(v_i,v_j)}^t\}_{(v_i,v_j)\in\mathcal{E}^t}$ in Eq. (7);
6.     Compute edge classification loss $\mathcal{L}_e$ based on $\{(\boldsymbol{e}_{(v_i,v_j)}^s, z_{(v_i,v_j)}^s)\}_{(v_i,v_j)\in\mathcal{E}^s}$ in Eq. (9);
7.     Compute supervised attention loss $\mathcal{L}_a$ based on $\{\{\alpha_{(v_i,v_j)}^{s(l)}\}_{l=1}^L, \{\alpha_{(v_j,v_i)}^{s(l)}\}_{l=1}^L, z_{(v_i,v_j)}^s\}_{(v_i,v_j)\in\mathcal{E}^s}$ in Eq. (12);
8.     Compute domain classification loss $\mathcal{L}_d$ based on $\{(\boldsymbol{e}_{(v_i,v_j)}^s, d_{(v_i,v_j)})\}_{(v_i,v_j)\in\mathcal{E}^s}$ and $\{(\boldsymbol{e}_{(v_i,v_j)}^t, d_{(v_i,v_j)})\}_{(v_i,v_j)\in\mathcal{E}^t}$ in Eq. (14);
9.     Backpropagate and update $\theta_h, \theta_y, \theta_z, \theta_d$ to optimize overall loss in Eq. (15);
10. **end while**
11. Apply optimized parameters $\theta_h^*$ to generate cross-network node and edge embeddings in Eq. (4) and (7);
12. Apply optimized parameters $\theta_z^*$ to predict target edge labels in Eq. (8).

back-propagation. Note that the GRL is not associated with any learnable parameters and does not perform during forward-propagation. However, during back-propagation, as shown in Fig. 5, the GRL reverses the partial derivative of the domain classification loss $\mathcal{L}_d$ w.r.t. the learnable parameters of the GNN encoder $f_h$ (i.e. $\frac{\partial \mathcal{L}_d}{\partial \theta_h}$) by multiplying it by $-\lambda$, where $\lambda > 0$ is the weight of the domain classification loss in the overall loss of DGASN. That is, after inserting the GRL, $\frac{\partial \mathcal{L}_d}{\partial \theta_h}$ is effectively replaced with $-\lambda\frac{\partial \mathcal{L}_d}{\partial \theta_h}$ during back-propagation. As a result, maximizing the domain classification loss $\mathcal{L}_d$ w.r.t. the GNN encoder $f_h$ (i.e. $\max_{\theta_h}\{\mathcal{L}_d\}$) can be optimized together with the minimization of $\mathcal{L}_d$ w.r.t. the domain discriminator $f_d$ (i.e. $\min_{\theta_d}\{\mathcal{L}_d\}$), during each back-propagation.

### E. Model Optimization

In the proposed DGASN, the GNN encoder $f_h$, node classifier $f_y$, edge classifier $f_z$ and domain discriminator $f_d$ are trained end-to-end by optimizing the overall minimax objective

function:

$$\mathcal{L} = \min_{\theta_h,\theta_y,\theta_z} \left\{ \mathcal{L}_e + \eta\mathcal{L}_n + \xi\mathcal{L}_a + \lambda\max_{\theta_d}\{-\mathcal{L}_d\} \right\} \quad (15)$$

where $\eta$, $\xi$ and $\lambda$ are all the trade-off hyper-parameters to balance the effect of different loss terms.

Algorithm 1 shows the training process of DGASN. The cross-network node embeddings are learned in Line 3, and the node classification loss is computed in Line 4. The cross-network edge embeddings are generated in Line 5 and the edge classification loss is computed in Line 6. The supervised attention loss is computed in Line 7. The domain classification loss is computed in Line 8. The proposed DGASN is trained end-to-end by optimizing the overall objective function in Line 9. After training convergence or reaching a maximum training epoch, the optimized learnable parameters are employed to generate cross-network node and edge embeddings in Line 11, and then predict the target edge labels in Line 12.

### F. Time Complexity

The time complexity of the GNN encoder constructed by multi-head GAT is $O\left(\left((|\mathcal{V}^s| + |\mathcal{V}^t|)\mathcal{W}\mathbb{d} + (|\mathcal{E}^s| + |\mathcal{E}^t|)\mathbb{d}\right)K\right)$, where $\mathbb{d}$ is the number of embedding dimensions of each head, $K$ is the number of attention heads, $|\mathcal{V}^s|$ and $|\mathcal{V}^t|$ are the number of nodes in $\mathcal{G}^s$ and $\mathcal{G}^t$ respectively, $|\mathcal{E}^s|$ and $|\mathcal{E}^t|$ are the number of edges in $\mathcal{G}^s$ and $\mathcal{G}^t$ respectively, $\mathcal{W}$ is the number of node attributes. The node classifier, edge classifier, and domain discriminator are all constructed as an MLP respectively. The time complexity of node classifier is linear to the number of nodes, and the time complexity of edge classifier and domain discriminator are both linear to the number of edges. Thus, the overall time complexity of DGASN is linear to number of nodes and edges in $\mathcal{G}^s$ and $\mathcal{G}^t$.

## IV. EXPERIMENTS

We conducted extensive experiments to investigate the following research questions (RQs):
- **RQ 1**: How does the proposed DGASN perform in the CNHHEC problem compared with the baselines?
- **RQ 2**: How do model variants affect the performance of DGASN?
- **RQ 3**: How do the hyper-parameters affect the performance of DGASN?

### A. Experimental Setup

*1) Datasets*

We adopted three real-world benchmark datasets [16] widely used for the cross-network classification tasks, including ACMv9, Citationv1 and DBLPv7. These three datasets inherently have varied data distributions. Each dataset is modeled as an undirected citation network, where each node represents a paper and each edge represents a citation relation between two papers. The sparse bag-of-words features extracted from the paper title are utilized as node attributes. Each paper can have multiple labels, belonging to some of the following five categories, including "Databases", "Artificial Intelligence", "Computer Vision", "Information Security", and "Networking", according to the relevant research topics. After removing self-loops, all edges in a network can be labeled as either homophilous or heterophilous, depending on the label agreement between two nodes on each edge. The statistics of the datasets are shown in Table II. Six CNHHEC tasks can be conducted among the three networks, i.e., C→A, D→A, A→C, D→C, A→D, C→D, where A, C, D denote ACMv9, Citationv1 and DBLPv7 respectively.

*2) Baselines*

The proposed DGASN was competed against two families of baselines, including 1) the GNNs designed for link prediction or noisy edge detection, and 2) the CNNC algorithms.

*Graph Neural Networks*: **VGAE** [44] employs a GCN encoder for node embedding learning. **AGE** [45] designs an adaptive encoder to iteratively strengthen the filtered features for better node embedding learning. Both VGAE and AGE use an inner product decoder for link prediction. **SuperGAT** [25] adopts GAT for node embedding learning and employs a self-supervised task of link prediction to directly supervise attention weights to distinguish connected node pairs from disconnected ones. **LAGCN** [11] learns node embeddings by SGC [27] and builds an edge classifier with MLP to classify edges into positive (homophilous) or negative (heterophilous). **RS-GNN** [12] learn an MLP-based link predictor based on node attributes and leverages the link predictor to down-weight noisy (heterophilous) edges.

*Cross-network Node Classification*: **CDNE** [16], **ACDNE** [19], **AdaGCN** [17], **UDAGCN** [20] and **ASN** [18] are state-of-the-art CNNC algorithms. CDNE employed two SAEs to learn node embeddings for the source network and the target network respectively. ACDNE adopts dual feature extractors to learn self-embeddings and neighbor-embeddings respectively.

TABLE II
STATISTICS OF THE REAL-WORLD NETWORKED DATASETS.

| Dataset | # Nodes | # Node Attributes | # Node Labels | # Edges | # Homophilous Edges | # Heterophilous Edges | # Self-loop Edges |
|---|---|---|---|---|---|---|---|
| ACMv9 | 9,360 | | | 15,602 | 13,883 | 1,673 | 46 |
| Citationv1 | 8,935 | 6,775 | 5 | 15,113 | 13,159 | 1,939 | 15 |
| DBLPv7 | 5,484 | | | 8,130 | 6,666 | 1,451 | 13 |

TABLE III
HYPER-PARAMETERS OF DGASN ON DIFFERENT TASKS.

| Task | # GAT Layers: $L$ | # Attention Heads: $K$ | # Embedding Dimensions of Each Head: $\mathbb{d}$ | Weight of $\mathcal{L}_n$: $\eta$ | Weight of $\mathcal{L}_a$: $\xi$ | Weight Decay |
|---|---|---|---|---|---|---|
| C→A | 8 | 8 | 64 | 1 | 1e-1 | 1e-3 |
| D→A | 3 | 8 | 64 | 1e-2 | 1e-1 | 1e-3 |
| A→C | 7 | 8 | 64 | 1 | 1e-3 | 5e-4 |
| D→C | 8 | 8 | 32 | 1 | 1e-4 | 1e-3 |
| A→D | 8 | 8 | 64 | 1 | 1e-2 | 1e-3 |
| C→D | 7 | 8 | 64 | 1 | 1e-1 | 5e-4 |

TABLE IV
AUC AND AP OF THE CNHHEC TASKS WITH THE CONCATENATE OPERATOR TO CONSTRUCT EDGE EMBEDDINGS. THE HIGHEST AUC AND AP AMONG ALL COMPARING ALGORITHMS ARE SHOWN IN BOLDFACE. (THE NUMBERS IN PARENTHESES ARE THE STANDARD DEVIATIONS OVER 5 RANDOM INITIALIZATIONS)

| Task | Metrics (%) | Graph Neural Networks | | | | | Cross-network Node Classification | | | | | DGASN (Ours) |
|---|---|---|---|---|---|---|---|---|---|---|---|---|
| | | VGAE | AGE | SuperGAT | LAGCN | RS-GNN | CDNE | ACDNE | ASN | UDAGCN | AdaGCN | |
| A→C | AUC | 50.5 (0.1) | 64.8 (0.6) | 59.3 (2.9) | 65.8 (0.9) | 67.7 (0.7) | 59.6 (0.2) | 68.0 (1.8) | 69.2 (3.2) | 68.8 (3.8) | 65.0 (4.1) | **76.8 (0.4)** |
| | AP | 13.6 (0.0) | 22.8 (0.3) | 16.8 (1.5) | 21.1 (0.6) | 21.8 (0.6) | 18.2 (0.1) | 25.1 (0.8) | 23.1 (1.9) | 24.9 (2.8) | 23.1 (3.1) | **30.9 (0.4)** |
| A→D | AUC | 51.1 (0.1) | 55.2 (0.9) | 55.4 (1.5) | 57.7 (1.3) | 63.5 (1.6) | 55.0 (0.1) | 60.04 (1.1) | 58.9 (1.6) | 63.8 (1.2) | 58.4 (3.1) | **67.5 (1.6)** |
| | AP | 19.3 (0.0) | 22.9 (0.3) | 20.9 (0.9) | 21.9 (1.2) | 26.4 (1.6) | 20.7 (1.7) | 25.0 (1.1) | 22.5 (0.9) | 28.3 (0.9) | 26.0 (1.2) | **30.7 (2.2)** |
| C→A | AUC | 53.0 (0.1) | 62.4 (0.3) | 61.2 (1.4) | 61.1 (0.9) | 62.3 (1.4) | 57.8 (0.1) | 64.7 (1.0) | 63.3 (0.9) | 64.8 (1.5) | 63.7 (1.4) | **70.9 (0.7)** |
| | AP | 13.2 (0.1) | 15.1 (0.4) | 15.2 (0.5) | 14.2 (0.5) | 15.9 (0.6) | 13.7 (0.1) | 18.1 (0.8) | 15.5 (0.5) | 16.7 (1.1) | 17.9 (0.8) | **23.8 (0.4)** |
| C→D | AUC | 56.2 (0.0) | 60.9 (0.7) | 57.6 (3.1) | 52.9 (1.3) | 60.6 (0.7) | 58.4 (0.2) | 61.8 (0.7) | 64.1 (0.2) | **66.1 (0.5)** | 62.0 (3.4) | 66.1 (0.6) |
| | AP | 20.8 (0.0) | 23.8 (0.5) | 21.9 (1.5) | 19.0 (0.4) | 24.4 (0.3) | 23.1 (0.3) | 26.5 (0.8) | 25.5 (0.7) | 28.6 (0.9) | 27.7 (4.2) | **29.6 (0.7)** |
| D→A | AUC | 53.3 (0.1) | 56.0 (1.4) | 56.6 (2.3) | 55.3 (1.2) | 57.3 (0.4) | 55.0 (0.5) | 57.1 (0.4) | 43.1 (2.1) | 59.6 (0.5) | 57.4 (2.7) | **65.9 (0.4)** |
| | AP | 13.0 (0.0) | 13.4 (0.5) | 13.6 (0.7) | 12.3 (0.4) | 13.7 (0.3) | 13.1 (0.3) | 15.0 (0.2) | 9.3 (0.3) | 15.2 (0.8) | 14.7 (0.9) | **18.0 (0.2)** |
| D→C | AUC | 54.8 (0.0) | 67.1 (0.6) | 58.8 (2.7) | 60.1 (1.1) | 63.4 (1.8) | 60.0 (0.5) | 66.0 (1.5) | 54.3 (7.4) | 60.1 (4.5) | 64.0 (7.0) | **73.1 (0.5)** |
| | AP | 15.5 (0.0) | 21.5 (0.7) | 16.8 (1.4) | 16.3 (0.6) | 19.5 (1.2) | 17.3 (0.2) | 22.3 (0.9) | 16.1 (2.4) | 17.7 (2.1) | 23.1 (4.0) | **26.2 (0.8)** |
| Average | AUC | 53.2 | 61.1 | 58.2 | 58.8 | 62.5 | 57.6 | 62.9 | 58.8 | 63.9 | 61.8 | **70.1** |
| | AP | 15.9 | 19.9 | 17.5 | 17.5 | 20.3 | 17.7 | 22.0 | 18.7 | 21.9 | 22.1 | **26.5** |

TABLE V
AUC AND AP OF CNHHEC WITH DIFFERENT OPERATORS TO CONSTRUCT EDGE EMBEDDINGS ON THE EXAMPLE TASK A→C. FOR EACH OPERATOR, THE HIGHEST AUC AND AP AMONG ALL COMPARING ALGORITHMS ARE SHOWN IN BOLDFACE.

| Operator | Metrics (%) | VGAE | AGE | SuperGAT | LAGCN | RS-GNN | CDNE | ACDNE | ASN | UDAGCN | AdaGCN | DGASN |
|---|---|---|---|---|---|---|---|---|---|---|---|---|
| Concatenate | AUC | 50.5 | 64.8 | 59.3 | 65.8 | 67.7 | 59.6 | 68.0 | 69.2 | 68.8 | 65.0 | **76.8** |
| | AP | 13.6 | 22.8 | 16.8 | 21.1 | 21.8 | 18.2 | 25.1 | 23.1 | 24.9 | 23.1 | **30.9** |
| Average | AUC | 49.7 | 66.2 | 60.9 | 65.3 | 67.3 | 61.9 | 68.0 | 69.2 | 64.8 | 64.7 | **72.4** |
| | AP | 12.6 | 24.6 | 18.2 | 20.9 | 20.8 | 19.7 | **26.2** | 22.9 | 21.2 | 20.9 | 25.0 |
| Hadamard | AUC | 53.5 | 67.0 | 46.5 | 61.6 | 67.1 | 54.2 | **69.5** | 66.8 | 64.4 | 64.4 | 52.7 |
| | AP | 15.0 | 24.1 | 12.1 | 20.5 | 21.5 | 16.4 | **26.3** | 21.3 | 21.9 | 21.3 | 16.4 |
| L1 | AUC | 53.3 | 62.8 | 45.6 | 64.6 | 59.0 | 54.2 | 61.9 | 60.8 | 54.8 | 64.1 | **68.4** |
| | AP | 14.4 | 19.6 | 11.6 | 20.8 | 18.6 | 15.2 | 20.5 | 18.8 | 17.2 | 23.4 | **27.2** |
| L2 | AUC | 53.4 | 59.8 | 54.1 | 60.6 | 57.3 | 55.0 | 60.9 | 60.7 | 54.3 | **61.3** | 53.9 |
| | AP | 14.3 | 18.5 | 15.2 | 17.8 | 17.6 | 15.7 | 19.9 | 18.6 | 16.3 | **21.5** | 14.8 |

AdaGCN, UDAGCN and ASN utilize GCN or GCN variants to learn node embeddings. To mitigate domain discrepancy, CDNE utilizes MMD-based domain adaptation [35]. AdaGCN adopts the Wasserstein distance guided adversarial domain adaptation [36]. UDAGCN, ACDNE and ASN employ the GRL-based adversarial domain adaptation [26].

It is worth noting that the GNN baselines [11, 12, 25, 44, 45] were designed for a single-network scenario. To tailor them to CNHHEC, we integrated the source and target networks into a single large network with the first $|\mathcal{V}^s|$ nodes from the source and the last $|\mathcal{V}^t|$ nodes from the target, and then employed the single network as the input network to the GNNs to learn node embeddings. While for the CNNC baselines [17, 19, 20] inherently developed for cross-network scenario, the node embeddings across networks can be learned directly. Then, for all the GNN and CNNC baselines, the edge embeddings were constructed based on the embeddings of two nodes on each edge, by adopting the same operator in Eq. (7) as the proposed DGASN. Next, given the edge embeddings as the input, an MLP (with the same setting of edge classifier $f_z$ in DGASN) was adopted to build an edge classifier to train on the source labeled edges and then predict the target edge labels.

However, it is infeasible to compare the proposed DGASN with the early cross-network edge classification algorithms [38-42], since the definitions of edge features and edge labels in such works are totally different from ours.

*3) Implementation Details*

The proposed DGASN[1] was implemented in PyTorch 1.10.2 [46] and Deep Graph Library (DGL) 0.8.2 [47]. DGASN was trained by the Adam optimizer. Following [26], the learning rate was decayed as $\mu_p = \frac{\mu_0}{(1+10p)^{0.75}}$, where the initial learning rate $\mu_0$ was set to 0.001, the training progress $p$ was linearly increased from 0 to 1, and the domain adaptation weight $\lambda$ was progressively increased from 0 to 0.1 as $\left(\frac{2}{1+\exp(-10p)} - 1\right) \times 0.1$. Both node classifier $f_y$ and edge classifier $f_z$ were constructed by an MLP with one hidden layer, with the number of hidden dimensions as 32 and 128 respectively. The domain classifier $f_d$ is constructed by an MLP with two hidden layers, and the hidden dimensions were set to 128 and 32 at the first and second hidden layers respectively. The cost-sensitive parameter $\gamma$ in Eq. (11) was set to 5. The other hyper-parameters, including the number of GAT layers $L$, the number of attention heads in the GAT encoder $K$, the number of embedding dimensions of each attention head $\mathbb{d}$, the weights of node classification loss $\eta$ and supervised attention loss $\xi$, and the weight decay to prevent over-fitting, are specified for each task in Table III.

---

[1] Our code is released at https://github.com/Qqqq-shao/DGASN.

Following the literatures on link prediction or binary edge classification [43, 45], we adopted two common metrics, i.e., the Area Under the ROC Curve (AUC) and Average Precision (AP) to evaluate the performance of homophilous and heterophilous edge classification on the target network. Each comparing method was repeated five times with different random initializations, and the averaged AUC and AP scores are reported in Table IV.

*B. Performance Comparison (RQ1)*

From Table IV, we have three observations as follows:

Firstly, the proposed DGASN consistently outperforms all baselines by a large margin on six tasks. On average, DGASN improves over the second-best method, i.e., UDAGCN by an absolute 6.2% and 4.6% in terms of AUC and AP.

Secondly, one can observe that the GNNs fail to achieve satisfactory performance in CNHHEC. This is because AGE, VGAE and SuperGAT are inherently developed for link prediction rather than homophilous and heterophilous edge classification. Similar to the proposed DGASN, SuperGAT also applies direct supervision on graph attention learning. However, SuperGAT guides the attention weights to distinguish neighbors from non-neighbors, while our DGASN guides the attention weights to distinguish homophilous neighbors from heterophilous neighbors. In addition, although LAGCN and RS-GNN employ a link predictor to discriminate homophilous edges from heterophilous edges, they are inherently designed for a single-network scenario without considering the domain discrepancy across networks. Thus, they would fail to learn network-invariant embeddings to address the cross-network edge classification problem.

Thirdly, the state-of-the-art CNNC methods also fail to succeed in the CNHHEC problem. This might be due to two factors. On one hand, ACDNE, ASN, UDAGCN and AdaGCN all utilize the GNNs with fixed edge weights for node embedding learning, which fail to distinguish the neighbors connected by homophilous edges from those connected by heterophilous edges. While the proposed DGASN employs multi-head GAT to learn adaptive edge weights during neighborhood aggregation and further applies direct supervision on graph attention learning, which can effectively alleviate the negative effect of heterophilous edges during neighborhood aggregation, and consequently yielding more label-discriminative node and edge embeddings to distinguish homophilous edges from heterophilous edges. On the other hand, although the CNNC baselines can effectively reduce domain discrepancy to learn network-invariant node embeddings, they do not learn node embeddings and edge embeddings jointly, thus, they fail to guarantee network-invariant edge embeddings. While a key design of the proposed DGASN is to jointly train node embeddings and edge embeddings in an end-to-end manner, thus, yielding label-discriminative and network-invariant edge embeddings to effectively address the CNHHEC problem.

Next, we adopt different operators in Eq. (7) to construct different types of edge embeddings. As shown in Table V, the proposed DGASN achieves the best performance among all methods when the edge embeddings are constructed by the concatenate, average and L1 operators. In addition, ACDNE outperforms other methods when the Hadamard operator is adopted, AdaGCN performs the best when the L2 operator is employed. Besides, one can see that most methods yield their best results when the concatenate operator is adopted to construct edge embeddings.

*C. Ablation Study (RQ2)*

Next, we conduct extensive ablation studies to investigate the contribution of each loss in the proposed DGASN. Firstly, as shown in in Table VI, the model variants without either node classification loss or edge classification loss would perform significantly worse than DGASN. This reflects that it is indeed necessary to jointly train node embeddings and edge embeddings supervised by the node classification and edge classification losses together. On one hand, the node classification loss guides label-discriminative node embeddings, which is essential to distinguish heterophilous edges from homophilous edges. On the other hand, the edge classification loss directly guides label-discriminative edge embeddings, note that the edge embeddings are constructed

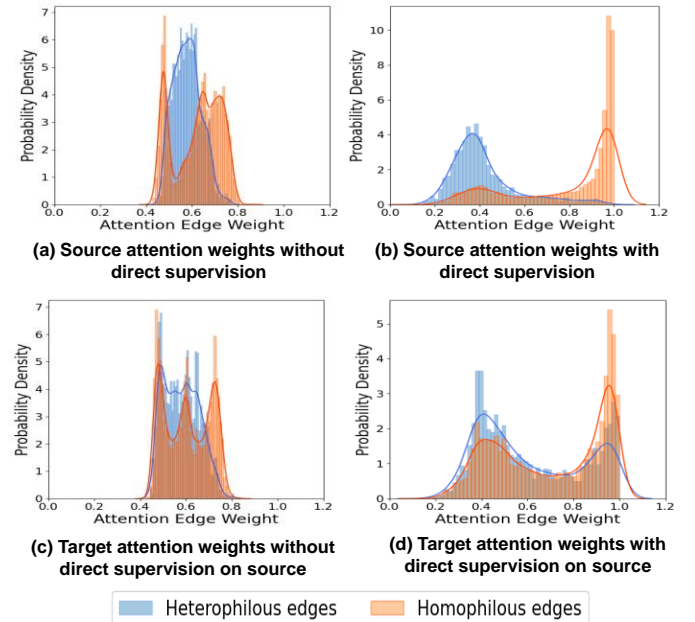

(a) Source attention weights without direct supervision
(b) Source attention weights with direct supervision
(c) Target attention weights without direct supervision on source
(d) Target attention weights with direct supervision on source

Fig. 6. Distributions of the source and target attention edge weights on the representative task C→D, when direct attention supervision is applied on the labeled source network or not (note that direct attention supervision is never applied on the unlabeled target network).

TABLE VI
AUC AND AP OF DGASN VARIANTS. THE HIGHEST AUC AND AP AMONG ALL MODEL VARIANTS ARE SHOWN IN BOLDFACE.

| Model Variants | Metrics (%) | A→C | A→D | C→A | C→D | D→A | D→C |
|---|---|---|---|---|---|---|---|
| DGASN | AUC | **76.8** | **67.5** | **70.9** | **66.1** | 65.9 | **73.1** |
|  | AP | **30.9** | **30.7** | **23.8** | **29.6** | **18.0** | **26.2** |
| w/o Node Classification Loss | AUC | 68.3 | 59.3 | 68.0 | 60.3 | **66.1** | 57.5 |
|  | AP | 24.5 | 24.5 | 19.3 | 21.6 | 17.9 | 13.8 |
| w/o Edge Classification Loss | AUC | 75.0 | 65.1 | 69.7 | 63.1 | 50 | 71.6 |
|  | AP | 29.8 | 30.3 | 22.1 | 27.4 | 10.8 | 24.0 |
| w/o Domain Classification Loss | AUC | 70.7 | 67.1 | 64.4 | 57.2 | 62.2 | 58.8 |
|  | AP | 25.3 | 29.1 | 17.9 | 24.6 | 17.7 | 20.4 |
| w/o Supervised Attention Loss | AUC | 76.2 | 65.8 | 68.3 | 64.3 | 64.7 | 72.9 |
|  | AP | 30.3 | 30.0 | 21.0 | 28.4 | 16.8 | 25.9 |

based on node embeddings, thus this in turn yields more label-discriminative node embeddings. In addition, without domain classification loss significantly degenerates the performance of DGASN. This reflects that reducing domain discrepancy is essential for CNHHEC.

Moreover, as shown in Table VI, without supervised attention loss would lead to significantly lower AUC and AP scores on all tasks. In addition, as shown in Fig. 6(b), with the supervised attention loss on the source network, the source homophilous edges would have larger attention weights than the source heterophilous edges. While without direct attention supervision on the source network, some source homophilous edges even have smaller attention weights than the source heterophilous edges, as shown in Fig. 6(a). Note that in DGASN, we only apply supervised attention loss on the fully labeled source network based on the observed source edge labels, while such supervised attention loss cannot be directly applied to the unlabeled target network. However, as shown in Fig. 6(d), with direct attention supervision on the source network, then for the unlabeled target network even without direct attention supervision, most target homophilous edges still possess larger attention weights than the target heterophilous edges. This is because by employing adversarial domain adaptation, the network-invariant edge embeddings can be learned by DGASN, i.e., the target network can have similar distributions of edge embeddings with that of the source network. The results on Fig. 6 and Table VI consistently verify the effectiveness of direct supervision on graph attention learning on discriminating homophilous edges from heterophilous edges across networks.

### D. Parameter Sensitivity (RQ3)

Next, we study the sensitivity of DGASN to the hyper-parameters $K, L, \mathbb{d}, \xi, \eta, \gamma$ on the representative task C→A.

The hyper-parameter $K$ represents the number of attention heads. As shown in Fig. 7(a), $K$=8 improves on $K$=1 by a large margin in terms of AUC and AP. This reflects that the multi-head attention mechanism is indeed beneficial for learning robust embeddings. While too many heads (i.e. $K$=16) would significantly degenerate the performance. Such performance drop might be due to two folds. On one hand, since different attention heads utilize unshared learnable parameters, too many heads would induce a lot of learnable parameters for attention learning, which increases the learning difficulty and easily causes over-fitting. On the other hand, in multi-head GAT, the final node embedding vector is generated by concatenating the embeddings learned by multiple heads with the dimension of $K\mathbb{d}^{(l)}$. Thus, a large number of heads $K$ results in too large dimensions for both node embeddings and edge embeddings, which again increases the difficulty for model learning.

The hyper-parameter $L$ denotes the number of GAT layers. As shown in Fig. 7(b), DGASN performs better by increasing the depth of GAT when $L \leq 8$. Note that each GAT layer only leverages the neighborhood information from one-hop neighbors. To aggregate the information from *L*-hop neighbors, it is required to stack *L* layers. It has been widely acknowledged that leveraging the information from high-order neighborhood is beneficial for node classification [16, 19, 20]. Such label-discriminative node embeddings to effectively separate different node classes can contribute to also label-discriminative edge embeddings to discriminate homophilous edges from heterophilous edges. However, stacking too many GNN layers (i.e. *L*=9) has been proven to easily cause over-smoothing [13, 48], i.e., making the node embeddings of different classes indistinguishable. Such over-smoothed node embeddings lead to indistinguishable edge embeddings between homophilous and heterophilous edges, consequently degrading the CNHHEC performance.

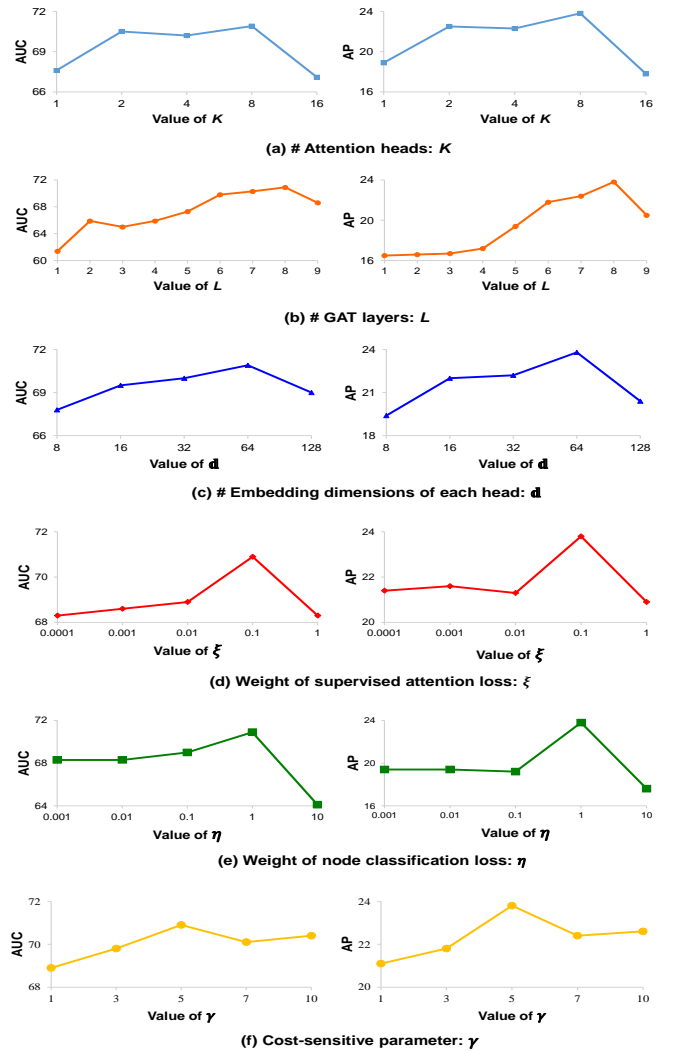

Fig. 7. Parameter sensitivity of DGASN on the representative task C→A.

The hyper-parameter $d$ is the number of embedding dimensions of each attention head. As shown in Fig. 7(c), both higher AUC and AP can be obtained, as $d$ increases when $d \in \{8, 16, 32, 64\}$. While when $d$ is further increased to 128, both AUC and AP exhibit a decreasing trend. Such performance degradation might be because too large embedding dimensions of node and edge embeddings would increase the number of learnable parameters and yield high difficulty in model learning.

The hyper-parameters $\xi$ and $\eta$ are the weight of supervised attention loss and node classification loss, respectively. As shown in Fig. 7(d) and 7(e), the performance of DGASN is sensitive to the values of $\xi$ and $\eta$. Specifically, $\xi = 0.1$ and $\eta = 1$ yield the best results on the task C→A.

The hyper-parameter $\gamma$ denotes the ratio of the penalty on the error of attention learning for heterophilous edges over that of homophilous edges. Specifically, $\gamma > 1$ means imposing larger penalty on the error of the heterophilous edges. As shown in Fig. 7(f), DGASN performs the best when $\gamma = 5$. In addition, $\gamma > 1$ all yields superior results than $\gamma = 1$. This is because for the homophilic graphs studied in our work, the number of heterophilous edges is much smaller than that of homophilous edges, making the supervised attention learning bias towards the homophilous edges. Setting $\gamma > 1$ makes the supervised attention learning focus more on the scarce heterophilous edges, which consequently yields better discrimination between heterophilous and homophilous edges.

## V. Conclusion

In this work, we make the pioneering attempt to study a novel CNHHEC problem. A novel framework named DGASN is proposed to effectively tackle the CNHHEC problem. DGASN adopts multi-head GAT as the GNN encoder, and employs a joint training strategy to train node embeddings and edge embeddings together, thus yielding informative embeddings to distinguish homophilous edges from heterophilous edges. In addition, in original GAT, the supervision on graph attention weights is limited and indirect. To improve the expressive power of graph attention weights in GAT, DGASN proposes to apply direct supervision on source attention weights, according to the observed edge labels in the source network. As a result, lower attention weights would be assigned to heterophilous edges so as to alleviate negative effect of the inter-class edges on neighborhood aggregation and yield more label-discriminative embeddings to separate nodes from different classes. Besides, DGASN employs the adversarial domain adaptation technique to learn network-invariant edge embeddings to facilitate knowledge transfer across networks. Extensive experiments on benchmark datasets demonstrate that the proposed DGASN can consistently gain superior results compared to the state-of-the-art GNNs and CNNC methods.

There are several directions for future research. Firstly, the proposed DGASN only aims to detect whether an existing edge in the target network is homophilous or heterophilous. It is interesting to develop new models to predict potential (or missing) homophilous and heterophilous edges between disconnected nodes in the target network. Secondly, this work focuses on the CNHHEC problem across undirected networks, how to address CNHHEC for directed networks remains an open challenging problem. Thirdly, recent single-network-based GNNs [9-12, 14, 15] showed that updating network topology based on the prediction of label agreement between node pairs can significantly improve the node classification performance. Thus, it is promising to make more exploration to jointly study the CNHHEC and CNNC problems. More research is needed to figure out how to take advantage of the predicted edges labels of CNHHEC to regularize the GNNs and improve the node classification performance in the target network.

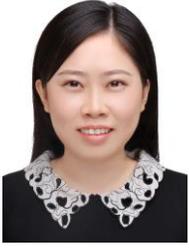
**Xiao Shen** received the double B.Sc. degrees in e-commerce engineering from Beijing University of Posts and Telecommunications and Queen Mary University of London in 2012, the M.Phil. degree in advanced computer science from University of Cambridge in 2013, and the Ph.D. degree in computer science from Hong Kong Polytechnic University, in 2019. She received the Hong Kong PhD Fellowship. She was a Postdoc Fellow at the Centre for Smart Health, Hong Kong Polytechnic University between 2019 and 2021. She is now an Associate Professor with the School of Computer Science and Technology, Hainan University, China. Her research interests include graph neural networks, graph contrastive learning, and cross-network classification. She has published in prestige international journals and conferences, including IEEE TKDE, TNNLS, TFS, TCyb, SIGIR, WWW and AAAI.

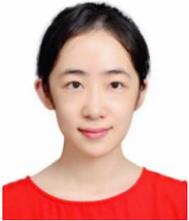
**Mengqiu Shao** received the bachelor's degree in computer science and technology from Hefei University of Technology, Anhui, China, in 2019. She is currently pursuing the master's degree in electronic information with the School of Computer Science and Technology, Hainan University. Her research interests include graph neural networks and cross-network classification.

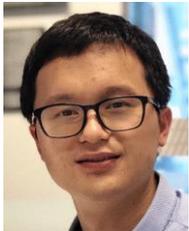
**Shirui Pan** (Member, IEEE) received the Ph.D. degree in information technology from the University of Technology Sydney, Ultimo, NSW, Australia. He is a Professor and an ARC Future Fellow with the School of Information and Communication Technology, Griffith University, Australia. Before joining Griffith in August, 2022, he was with the Faculty of Information Technology, Monash University between Feb 2019 and July 2022. His research interests include data science and AI. He has authored or coauthored more than 100 research papers in top-tier journals and conferences, including the TPAMI, TKDE, TNNLS, ICML, NeurIPS, KDD, AAAI, IJCAI, WWW, CVPR, and ICDM. He was the recipient of the Best Student Paper Award of IEEE ICDM 2020. He is recognized as one of the AI 2000 AAAI/IJCAI Most Influential Scholars in Australia, in 2021.

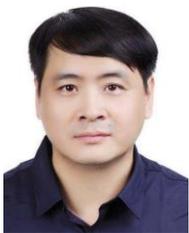
**Laurence T. Yang** (Fellow, IEEE) received the B.E. degree in computer science and technology and the B.Sc. degree in applied physics both from Tsinghua University, China, and the Ph.D. degree in computer science from the University of Victoria, Canada. He is a Professor with the School of Computer Science and Technology, Hainan University, China, and with the Department of Computer Science, St. Francis Xavier University, Canada. His research interests include cyber-physical-social systems, parallel and distributed computing, embedded and ubiquitous/pervasive computing, and big data. His research has been supported by the National Sciences and Engineering Research Council, Canada, and the Canada Foundation for Innovation.

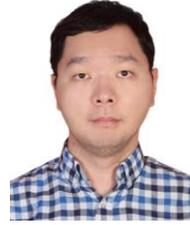
**Xi Zhou** received the B.Sc. degree in Biotechnology from Fudan University in 2008, and the Ph.D. degree in Bioinformatics from Zhejiang University in 2013. He was a Research Fellow at the Centre for Smart Health, Hong Kong Polytechnic University. He is now an Assistant Professor with College of Tropical Crops, Hainan University, China. His research interests include graph neural networks and protein-protein interaction prediction.